\documentclass{article}
\usepackage{spconf,amsmath,graphicx}
\usepackage{multirow}
\usepackage{booktabs}
\usepackage{cite}
\usepackage{color}
\usepackage{enumitem}
\setlist{nosep, leftmargin=14pt}
\usepackage{mathrsfs}
\usepackage{amssymb}
\usepackage{mwe} 
\usepackage[colorlinks=true, linkcolor=red, citecolor=green]{hyperref}

\newcommand{\ie}{i.e., }
\newcommand{\eg}{e.g., }
\title{Joint Rigid Motion Correction and Sparse-View CT\\ via Self-Calibrating Neural Field}

\begin{document}
\ninept
%
%
%
%

\name{Qing Wu$^{\dagger}$ \qquad Xin Li$^{\dagger}$ \qquad Hongjiang Wei$^{\ddagger}$ \qquad Jingyi Yu$^{\dagger}$ \qquad Yuyao Zhang$^{\dagger,\star}$}
\address{$^{\dagger}$ School of Information Science and Technology, ShanghaiTech University, Shanghai, China\\
    $^{\ddagger}$School of Biomedical Engineering, Shanghai Jiao Tong University, Shanghai, China}
\maketitle
\begin{abstract}
Neural Radiance Field (NeRF) has widely received attention in Sparse-View Computed Tomography (SVCT) reconstruction tasks as a self-supervised deep learning framework. NeRF-based SVCT methods represent the desired CT image as a continuous function of spatial coordinates and train a Multi-Layer Perceptron (MLP) to learn the function by minimizing loss on the SV sinogram. Benefiting from the continuous representation provided by NeRF, the high-quality CT image can be reconstructed. However, existing NeRF-based SVCT methods strictly suppose there is completely no relative motion during the CT acquisition because they require \textit{accurate} projection poses to model the X-rays that scan the SV sinogram. Therefore, these methods suffer from severe performance drops for real SVCT imaging with motion. In this work, we propose a self-calibrating neural field to recover the artifacts-free image from the rigid motion-corrupted SV sinogram without using any external data. Specifically, we parametrize the inaccurate projection poses caused by rigid motion as trainable variables and then jointly optimize these pose variables and the MLP. We conduct numerical experiments on a public CT image dataset. The results indicate our model significantly outperforms two representative NeRF-based methods for SVCT reconstruction tasks with four different levels of rigid motion.
\end{abstract}
\section{Introduction}
\label{sec:intro}
\par Sparse-View Computed Tomography (SVCT) can significantly reduce the radiation dose and shorten the scanning time by decreasing the number of radiation views. However, the insufficient projection measurement (\ie SV sinogram) in SVCT will suffer from severe streaking artifacts if applying conventional analytical reconstruction algorithms such as Filtered Back-Projection (FBP) \cite{fbp}, which significantly degrades image quality. 
\par Recently, several self-supervised SVCT methods \cite{tancik2020fourier,shen2022nerp,wu2022self,zang2021intratomo,zha2022naf} based on Neural Radiance Field (NeRF) \cite{mildenhall2020nerf} have been emerged. Different from supervised deep learning models \cite{zhang2018sparse, FBPConvNet}, these NeRF-based methods can recover the high-quality CT image from the SV sinogram without using any external data. Specifically, NeRF-based methods first represent the unknown CT image as a continuous function that maps coordinates to intensities and then train a Multi-Layer Perceptron (MLP) to learn the function by minimizing prediction errors on the SV sinogram. Benefiting from the implicit continuous prior imposed by the function and the neural network prior \cite{ulyanov2018deep, xu2019frequency, rahaman2019spectral}, the function can be approximated well, and thus the desired high-quality CT image will be reconstructed.
\par Existing NeRF-based SVCT methods \cite{tancik2020fourier,shen2022nerp,wu2022self,zang2021intratomo,zha2022naf} suppose there is completely no relative motion during the CT acquisition process, which benefits that the \textit{accurate} projection poses can be accessible for modeling the X-rays that scan the SV sinogram. However, relative motion, especially rigid motion, is common and even inevitable \cite{ko2021rigid, sun2021motion, kim2015rigid, zhang2010correction} during the real CT acquisition process due to various factors (\eg imaging subject's movement and CT scanner's system error, etc.). Therefore, this overly strict assumption will result in severe model performance drops in real SVCT reconstruction with motion.
\par In this paper, we propose a self-calibrating neural field that can reconstruct the artifacts-free image from the rigid motion-corrupted SV measurement without involving any external data. Like the previous works \cite{tancik2020fourier,shen2022nerp,wu2022self,zang2021intratomo,zha2022naf}, our proposed model is also based on NeRF's \cite{wang2021nerf} framework (\ie using an MLP to learn the function of desired CT image through minimizing loss on the SV sinogram). The major novelty of our model is that it extra models the rigid motion in the CT acquisition process and thus can produce robust and excellent results for SVCT imaging with rigid motion. More specifically, we first parameterize inaccurate projection poses caused by rigid motion as three trainable variables (a rotation angle and two translation offsets). Then, we jointly optimize these pose variables and the MLP representation. After the poses calibration and MLP optimization, the final high-quality CT image can be reconstructed. We conduct numerical experiments on a public COVID-19 CT image dataset \cite{shakouri2021covid19}. Experimental results shows that the proposed model significantly outperforms the two representative NeRF-based methods \cite{tancik2020fourier,shen2022nerp} for SVCT reconstruction tasks with four different levels of rigid motion. We also perform ablation study for the pose correction module in our model. The results confirm its effectiveness.

\begin{figure*}[t]
    \centering
    \includegraphics[width=0.8\textwidth]{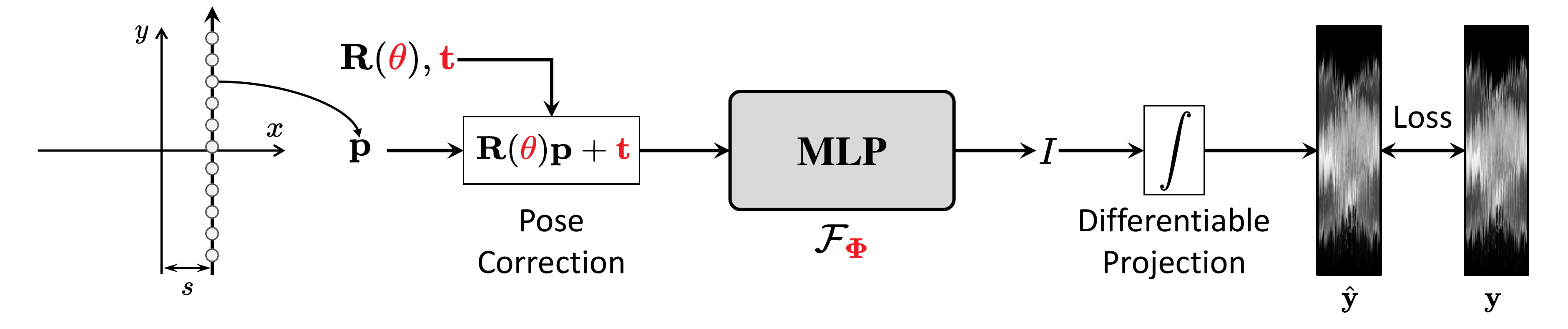}
    \caption{Pipeline of our proposed method for SVCT with rigid motion. Here the trainable parameters are highlighted in \textcolor{red}{red}.}
    \label{fig:method}
\end{figure*}
\section{Background}
\par Formally, NeRF-based SVCT methods \cite{tancik2020fourier, shen2022nerp, wu2022self, zang2021intratomo,zha2022naf} represent the unknown and high-quality CT image $\mathbf{x}\in \mathbb{R}^{N\times N}$ as a continuous function:
\begin{equation}
    \mathcal{M}: \mathbf{p} \rightarrow I,
\end{equation}
where $\mathbf{p}=(x,y)\in\mathbb{R}^2$ is any spatial coordinate in a normalized 2D Cartesian coordinates $[-1, 1]\times[-1, 1]$ and $I\in\mathbb{R}$ denotes the corresponding intensity value in the CT image $\mathbf{x}$.
\par Given the acquired SV sinogram $\mathbf{y}\in\mathbb{R}^{M\times N}$, where $M$ and $N$ denote the number of projections and X-rays per projection, respectively. NeRF-based methods leverage an MLP $\mathcal{F}_\mathbf{\Phi}$ to learn the continuous function $\mathcal{M}$ by optimizing the objective as below:
\begin{equation}
    \mathbf{\Phi}^* = \mathop{\arg\min}\limits_{\mathbf{\Phi}}\ \mathcal{L}(\mathbf{y}, \mathbf{A}\mathcal{F}_{\mathbf{\Phi}}),
\end{equation}
where $\mathbf{A}\in \mathbb{R}^{N\times M}$ represents a differentiable projection operator (\eg Radon transform for parallel X-ray beam CT acquisition) and $\mathcal{L}$ is the similarity metric used for measuring data discrepancy between the generated SV measurement $\mathbf{A}\mathcal{F}_{\mathbf{\Phi}}$ and the real SV sinogram $\mathbf{y}$. The function $\mathcal{M}$ can be approximated well benefiting from the implicit continuous prior provided by the function and neural network's learning bias toward low-frequency information \cite{ulyanov2018deep, xu2019frequency, rahaman2019spectral}. Once the optimization is converged, the high-quality CT image $\mathbf{x}$ can be reconstructed by feeding all the spatial coordinates $\mathbf{p}$ into the MLP $\mathcal{F}_{\mathbf{\Phi}^*}$ to predict the corresponding intensity values $I$. 
\section{Proposed Model}
\label{sec:method}
\par Although existing NeRF-based SVCT reconstruction methods \cite{tancik2020fourier, shen2022nerp, wu2022self, zang2021intratomo, zha2022naf} have shown great potential, there is still a major limitation: The \textit{accurate} projection poses have to be available for modeling the X-ray that scan SV sinogram. However, inevitable relative motion during the CT acquisition always results in the inaccurate projection poses. Inspired by NeRF$--$ \cite{wang2021nerf}, we parametrize all the inaccurate projection poses as trainable variables, and jointly optimize these variables and the MLP. Therefore, our proposed method is able to recover the high-quality CT image from the rigid-motion-corrupted SV sinogram. In this section, our model is introduced in detail.
\subsection{Projection Pose Parameterization}
\par In the proposed model, we make two basic assumptions: (1) The type of motion is rigid (\ie $\text{degree of freedom}=3$ for 2D parallel X-ray beam CT); (2) No motion among X-rays from the same projection acquisition. This assumption is reasonable because the single projection is very fast (\eg it only takes about 0.3 seconds by using multi-slice CT \cite{ko2021rigid}). Based on these two assumptions, the X-rays from the same projection acquisition thus can share the same pose parameters. In particular, we leverage a rotation matrix $\mathbf{R}(\theta_i)\in \mathbb{SO}(2)$ ($\theta_i$ denotes the projection angle) and a translation vector $\mathbf{t}_i\in \mathbb{R}^2$ to parametrize the pose of the $i$-th projection acquisition as below:
\begin{equation}
    \mathcal{\mathbf{R}}(\theta_i)=\left[\begin{array}{rr}
    \cos{\theta_i} & -\sin{\theta_i}\\ 
      \sin{\theta_i}&\cos{\theta_i}
\end{array}\right], \quad    \mathbf{t}_i = \left[\begin{array}{rr}
    t_x^{i} & t_y^{i} 
    \end{array}\right]^\mathsf{T},
\end{equation}
\par Here we directly set the two elements $t_x^{i}$ and $t_y^{i}$ in the vector $\mathbf{t}_i$ as trainable variables for correcting translation motion because $\mathbf{t}_i$ is defined in Euclidean space $\mathbb{R}^2$, while we optimize the projection angle $\theta_i$ in the matrix $\mathbf{R}(\theta_i)$ for correcting rotation motion since $\mathbf{R}(\theta_i)$ is defined in $\mathbb{SO}(2)$ space.
\subsection{Jointing Pose Correction and MLP Optimization}
\par Fig. \ref{fig:method} illustrates the pipeline of our proposed model. To generate any projection value $\mathbf{y}(\theta, s)$ of the acquired rigid motion-corrupted SV sinogram, we first build a X-ray $L_{s}: x=s$ in a standard space and sample intensity coordinates $\mathbf{p}$ along the X-ray. Then, we transform these sampled coordinates $\mathbf{p}$ into a real physic space by using pose correction, which is expressed as below:
\begin{equation}
    \mathbf{p}_\mathsf{real} = \mathbf{R}(\theta)\mathbf{p}+\mathbf{t},
\end{equation}
\par The MLP $\mathcal{F}_{\mathbf{\Phi}}$ takes these intensity coordinates $\mathbf{p}_\mathsf{real}$ in the real physic space as input and predicts the corresponding intensity values $I$. Then the estimated projection value $\hat{\mathbf{y}}(\theta, \rho)$ are generated by using differentiable line integral projection operator. Finally, we jointly optimize the pose variables $\{\mathbf{\theta},\mathbf{t}\}$ and the MLP's parameters $\mathbf{\Phi}$ by using gradient descent back-propagation algorithm to minimize the prediction error on the projection domain. Mathematically, the objective is expressed as below:
\begin{align}
    \mathbf{\Phi}^*, \theta^*, \mathbf{t}^* & = \mathop{\arg\min}\limits_{\mathbf{\Phi},\theta,\mathbf{t}}\ \mathcal{L}(\mathbf{y}, \hat{\mathbf{y}}),\\
    \hat{\mathbf{y}}(\theta, s) &= \sum_{\mathbf{p}\in{L_{s}}}{\mathcal{F}_{\mathbf{\Phi}}\left(\mathbf{R}\left(\theta\right)\mathbf{p}+\mathbf{t} \right)},
\end{align}
where $\hat{\mathbf{y}}\in\mathbb{R}^{M\times N}$ is estimated SV sinogram and the loss function $\mathcal{L}$ is implemented by $\ell_1$ norm. Moreover, all the elements in the vector $\mathbf{t}$ are initialed as 0 because any prior knowledge of the translation motion is assumed not to exist.
\begin{figure}[t]
    \centering
    \includegraphics[width=0.48\textwidth]{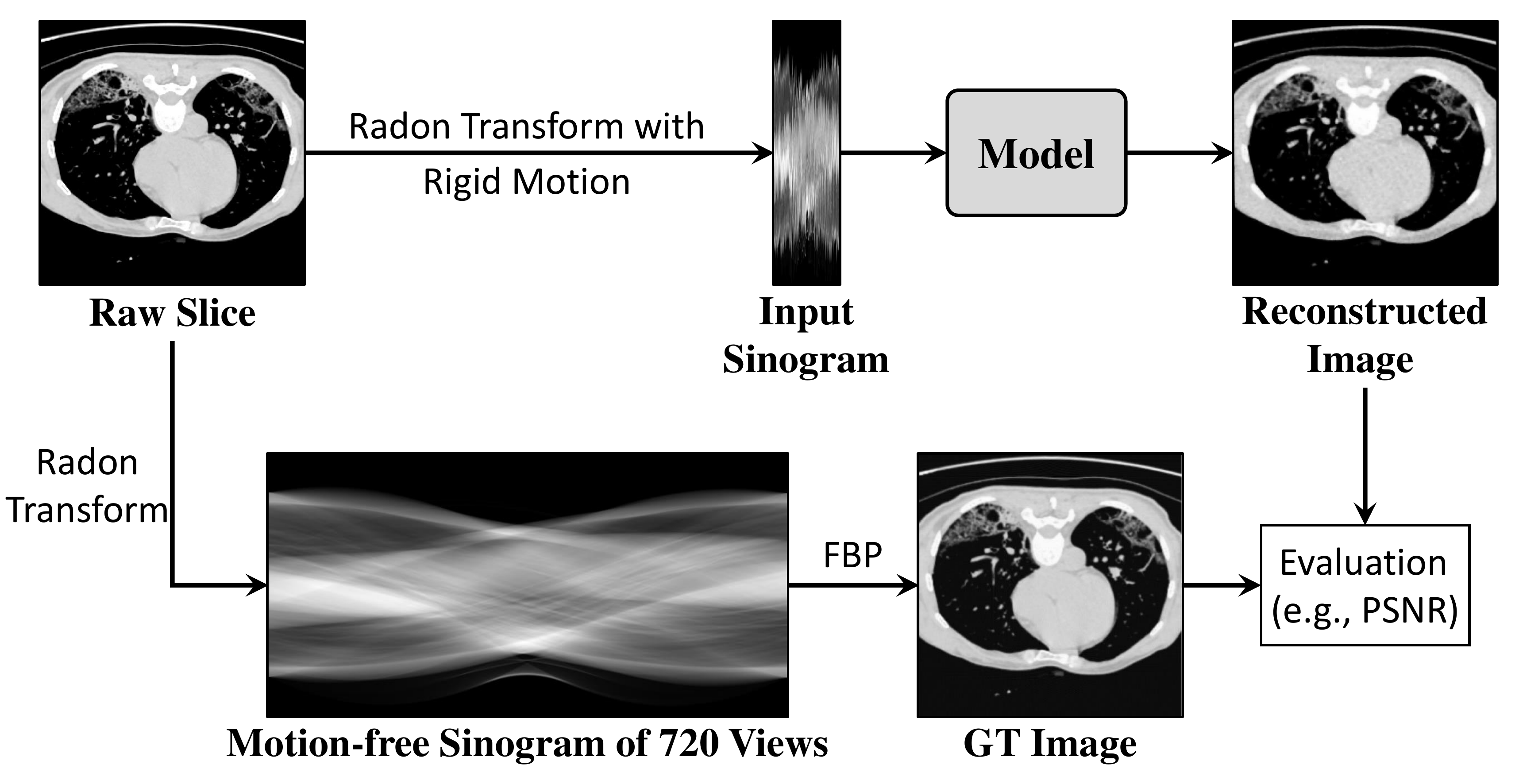}
    \caption{Workflow of our data processing.}
    \label{fig:data}
\end{figure}
\begin{figure*}[t]
    \centering
    \includegraphics[width=\textwidth]{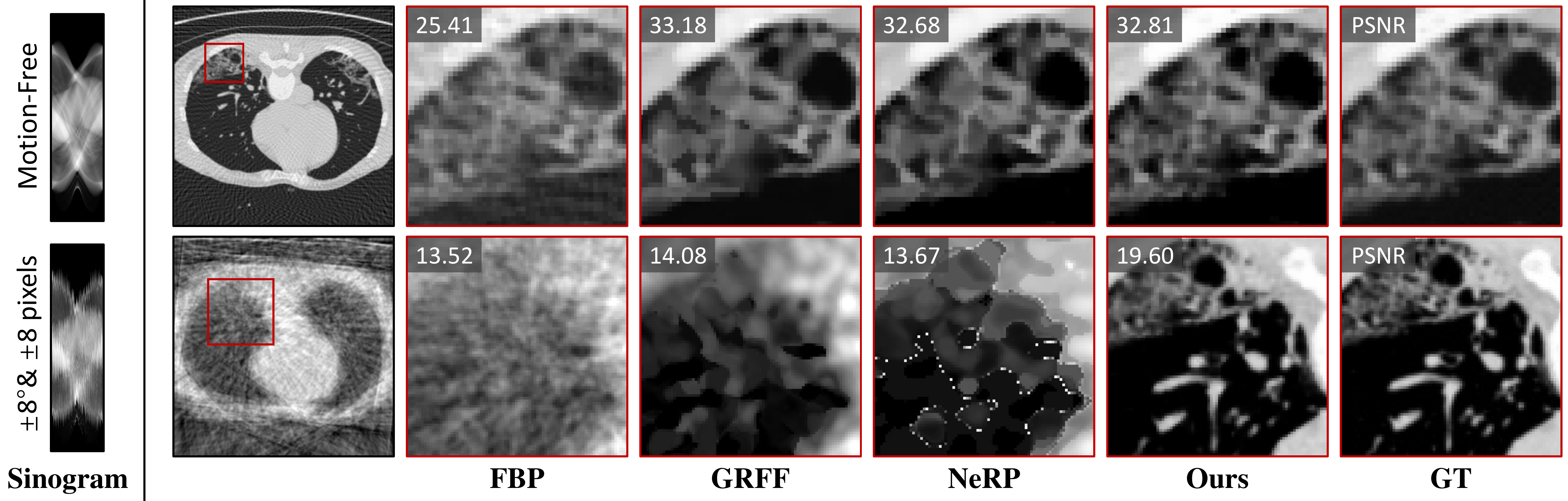}
    \caption{Qualitative results of all the compared methods on a sample of the COVID-19 dataset \cite{shakouri2021covid19} for SVCT without motion and with moderate rigid motion ($\pm8^\circ\&\pm8\ \mathsf{pixels}$).}
    \label{fig:com}
\end{figure*}
\begin{table*}[t]
    \centering
    \resizebox{0.9\textwidth}{!}{
    \begin{tabular}{c|c|c|c|c}
       \textbf{Motion Settings}  &\textbf{FBP} \cite{fbp} &  \textbf{GRFF} \cite{tancik2020fourier}&  \textbf{NeRP} \cite{shen2022nerp}&  \textbf{Ours}\\ \hline
        Motion-Free & $25.57/0.6334/0.2327$& $33.42/0.9524/0.0146$& $32.93/0.9488/0.0177$& $32.81/0.9437/0.0225$\\ \hline
        $\pm2^\circ\&\pm2\ \mathsf{pixels}$ & $18.47/0.3202/0.4295$& $19.72/0.3963/0.2875$& $19.29/0.3795/0.3273$& $24.17/0.8694/0.0276$\\
        $\pm8^\circ\&\pm8\ \mathsf{pixels}$ & $13.43/0.1081/0.5755$& $14.00/0.1529/0.5119$& $13.70/0.1589/0.5552$& $19.60/0.7564/0.0387$\\
        $\pm16^\circ\&\pm16\ \mathsf{pixels}$ & $11.21/0.0533/0.6319$& $11.60/0.0990/0.5776$& $11.50/0.1103/0.5987$& $16.85/0.6339/0.0570$\\ \hline
        \textbf{Mean} & $17.17/0.2788/0.4674$& $19.69/0.4001/0.3479$& $19.35/0.3994/0.3747$& $23.36/0.8008/0.0365$ \\
    \end{tabular}}
    \caption{Quantitative results (PSNR/SSIM/LPIPS) of all the compared methods on the COVID-19 dataset \cite{shakouri2021covid19} for SVCT with four levels of rigid motion. The higher PSNR and SSIM denote the better performance, while the lower LPIPS represent the better performance.}
    \label{tab:com}
\end{table*}
\subsection{Implementation Details}
\par In order to improve the neural network's learning ability to high-frequency information, we combine hash encoding \cite{muller2022instant} with two fully-connected layers to implement the MLP $\mathcal{F}_{\mathbf{\Phi}}$. Different from frequency encoding strategies (\eg Fourier encoding \cite{tancik2020fourier}), the hash encoding \cite{muller2022instant} is adaptive and thus can provide a better learning performance. For the model training, we utilize the Adam optimizer \cite{Kingma2015AdamAM} and its hyper-parameters are set as default. The learning rate is from 1e-3 and decays by a factor of 0.5 per 500 epochs. The total training epochs is 5000. 
\section{Numerical experiments}
\label{sec:exp}
\subsection{Experimental Settings}
\paragraph*{Dataset \& Pre-processing} All the numerical experiments in this paper are performed based on COVID-19 dataset \cite{shakouri2021covid19} that consists of 3D CT volumes from 1000+ patients with COVID-19 infections. We extract 2D slices of 256$\times$256 size from two 3D volumes in the COVID-19 dataset as experimental data. Fig. \ref{fig:data} shows the pipeline of our data processing. On the one hand, we build motion-free sinograms of 720 views by performing radon transformation on the raw slices and then generate GT images by FBP \cite{fbp}. The resulting GT images are used for model evaluation. On the other hand, we conduct radon transformation with rigid motion on the raw slices to generate motion-corrupted sinograms of 90 views. More specifically, for each projection simulation, the raw slices are translated by $t_x$ and $t_y$ $\mathsf{pixels}$ along $\mathsf{X}$ and $\mathsf{Y}$ axes respectively and then are rotated by $\alpha^\circ$ around the origin. The three independent motion parameters for each projection are sampled from the Uniform distribution $\mathcal{U}(-k, k)$. We set $k=\{0, 2, 8, 16\}$ to simulate a motion-free setting and three different levels of rigid motion. The generated sinograms are used for input data. Again, we would like to emphasize that all the compared methods directly reconstruct the corresponding CT images from the simulated SV sinograms.
\paragraph*{Compared Methods} We compare our proposed model with three SVCT methods: (1) FBP \cite{fbp}, a classical filtered back-projection CT imaging algorithm; (2) GRFF \cite{tancik2020fourier}, an earliest NeRF-based SVCT reconstruction method; (3) NeRP \cite{shen2022nerp}, a recent NeRF-based SVCT reconstruction method with prior embedding. Here FBP \cite{fbp} is based on the scikit-image library \cite{van2014scikit} of Python, while GRFF \cite{tancik2020fourier} and NeRP \cite{shen2022nerp} are implemented following the original papers.
\begin{figure}[t]
    \centering
    \includegraphics[width=0.35\textwidth]{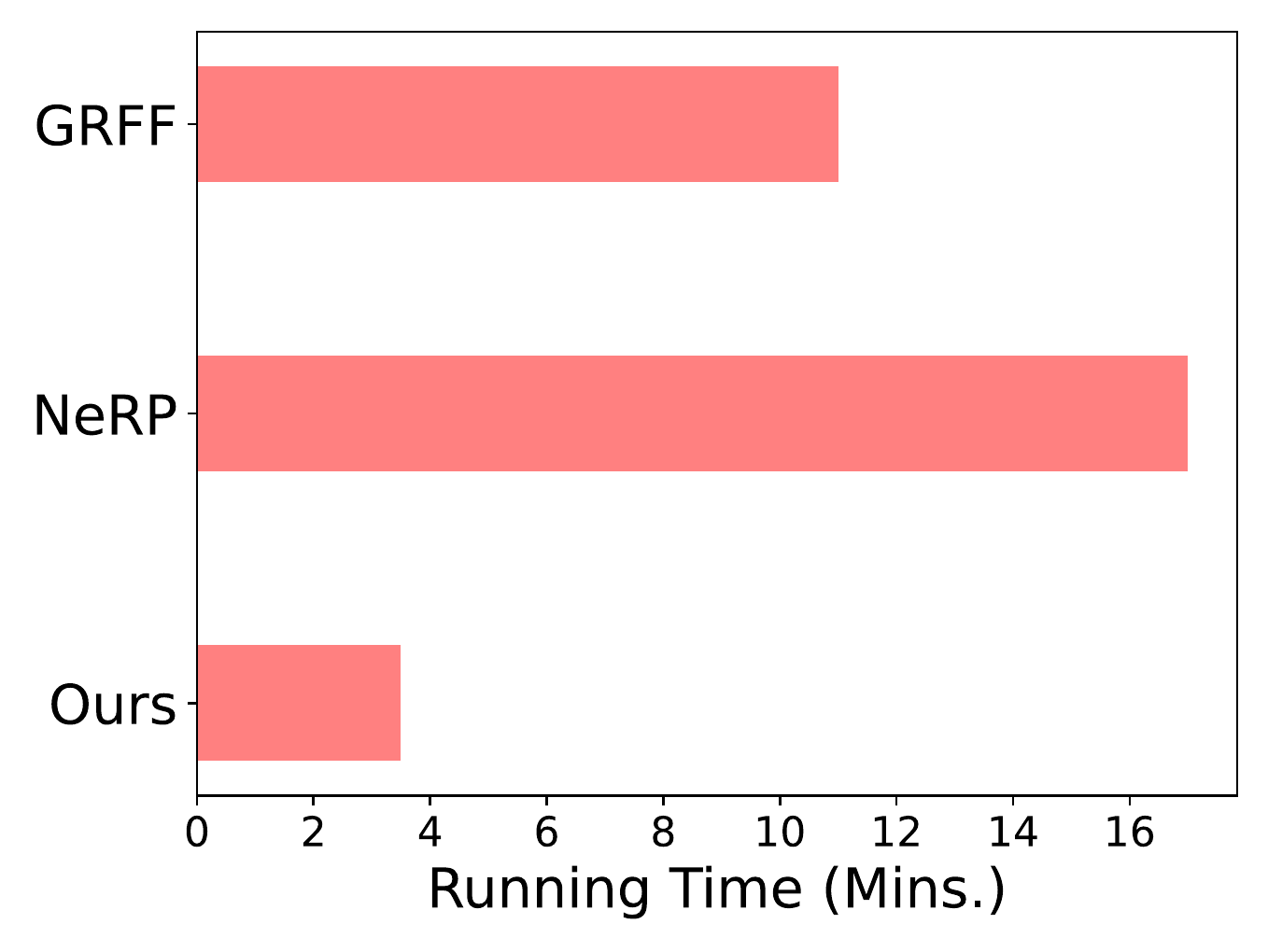}
    \caption{Average running time of GRFF \cite{tancik2020fourier}, NeRP \cite{shen2022nerp}, and our model for all the samples of the COVID-19 dataset.}
    \label{fig:running_time}
\end{figure}
\paragraph*{Evaluation Metrics} PNSR and SSIM \cite{ssim}, two well-known objective image quality metrics for low level vision tasks, are employed for quantitative evaluation. We also include LPIPS \cite{lpips}, a recent deep learning-based evaluation metric.
\subsection{Comparison with Other Methods}
\par We compare the proposed model with three baselines on the COVID-19 dataset \cite{shakouri2021covid19} for SVCT reconstruction with four motion settings. Table \ref{tab:com} shows the quantitative results and it indicates that the reconstruction capacity of all the compared models decreases with increasing movement range. For motion-free setting, the three NeRF-based methods (GRFF \cite{tancik2020fourier}, NeRP \cite{shen2022nerp}, and our model) all produce significantly improvements compared to FBP \cite{fbp}. While for the three levels of rigid motion, our proposed model is far better than the three baselines (FBP \cite{fbp}, GRFF \cite{tancik2020fourier}, and NeRP \cite{shen2022nerp}). For example, PSNR respectively improve 6.17 dB, 5.6 dB, and 5.9 dB for the moderate rigid motion ($\pm8^\circ\&\pm8\ \mathsf{pixels}$). Fig. \ref{fig:com} shows the qualitative results. For the motion-free setting, the three NeRF-based methods greatly improve the resulting image quality compared with FBP \cite{fbp}. And compared with GRFF \cite{tancik2020fourier} and NeRP \cite{shen2022nerp}, the result of our model has clearer and sharper image details. However, GRFF \cite{tancik2020fourier} and NeRP \cite{shen2022nerp} almost are failed for the moderate rigid motion ($\pm8^\circ\&\pm8\ \mathsf{pixels}$). This is mainly because they do not model the motion in the CT acquisition. In comparison, our method still produces an excellent result that is very close to GT image. Moreover, Fig. \ref{fig:running_time} shows the average running time of GRFF \cite{tancik2020fourier}, NeRP \cite{shen2022nerp}, and our model for all the samples of the COVID-19 dataset. Our model has a better performance on time consumption benefiting from the adaptive hash encoding module \cite{muller2022instant}. The time efficiency of our model is $3\times$ better to GRFF \cite{tancik2020fourier} and about $5\times$ better to NeRP \cite{shen2022nerp}.
\subsection{Effectiveness of Poses Correction}
\par
\begin{figure}[t]
    \centering
    \includegraphics[width=0.48\textwidth]{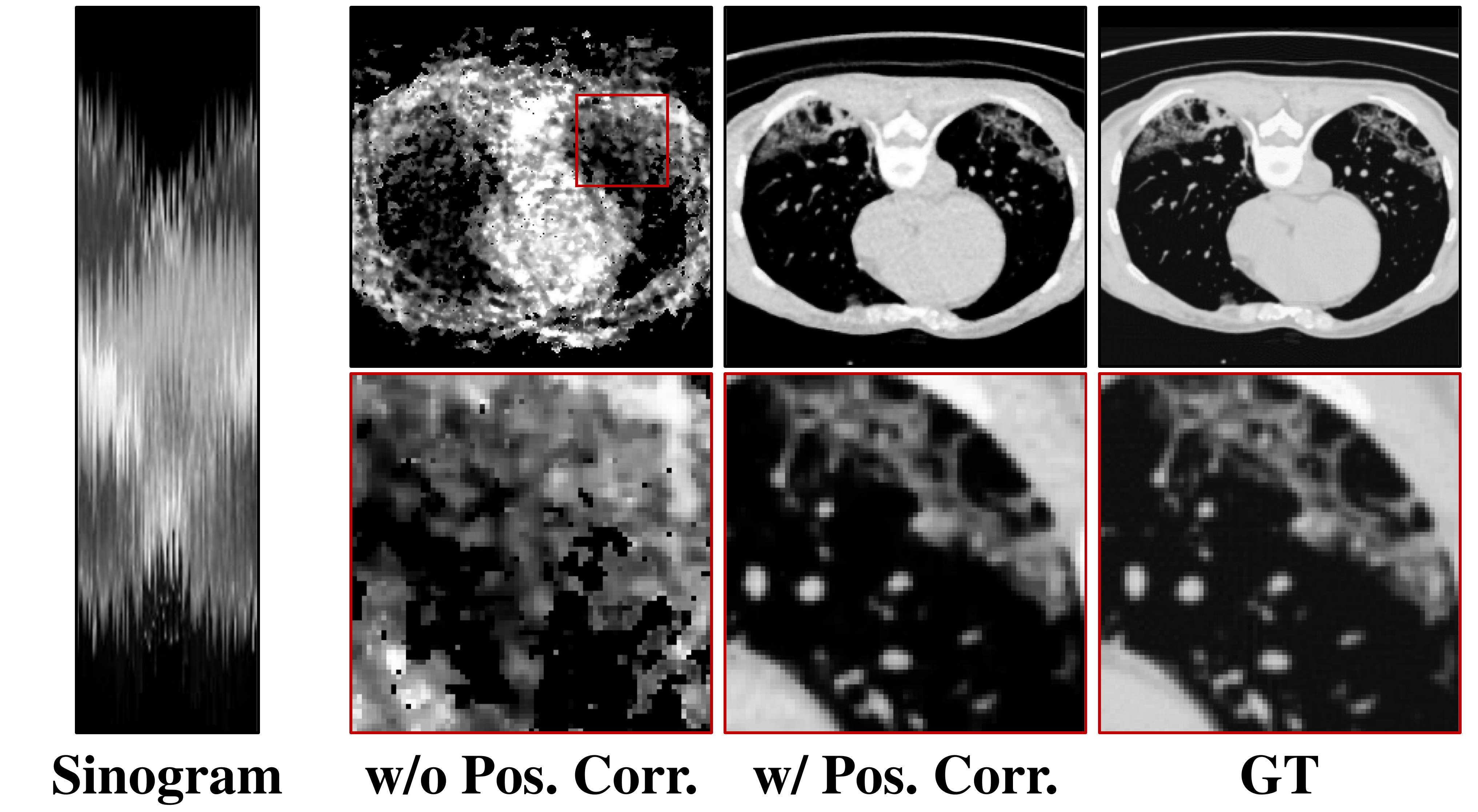}
    \caption{Qualitative results of our model without and with poses correction on a sample of the COVID-19 dataset \cite{shakouri2021covid19} for SVCT with severe rigid motion ($\pm16^\circ\&\pm16\ \mathsf{pixels}$).}
    \label{fig:pose_com}
\end{figure}

\begin{figure}[t]
    \centering
    \includegraphics[width=0.48\textwidth]{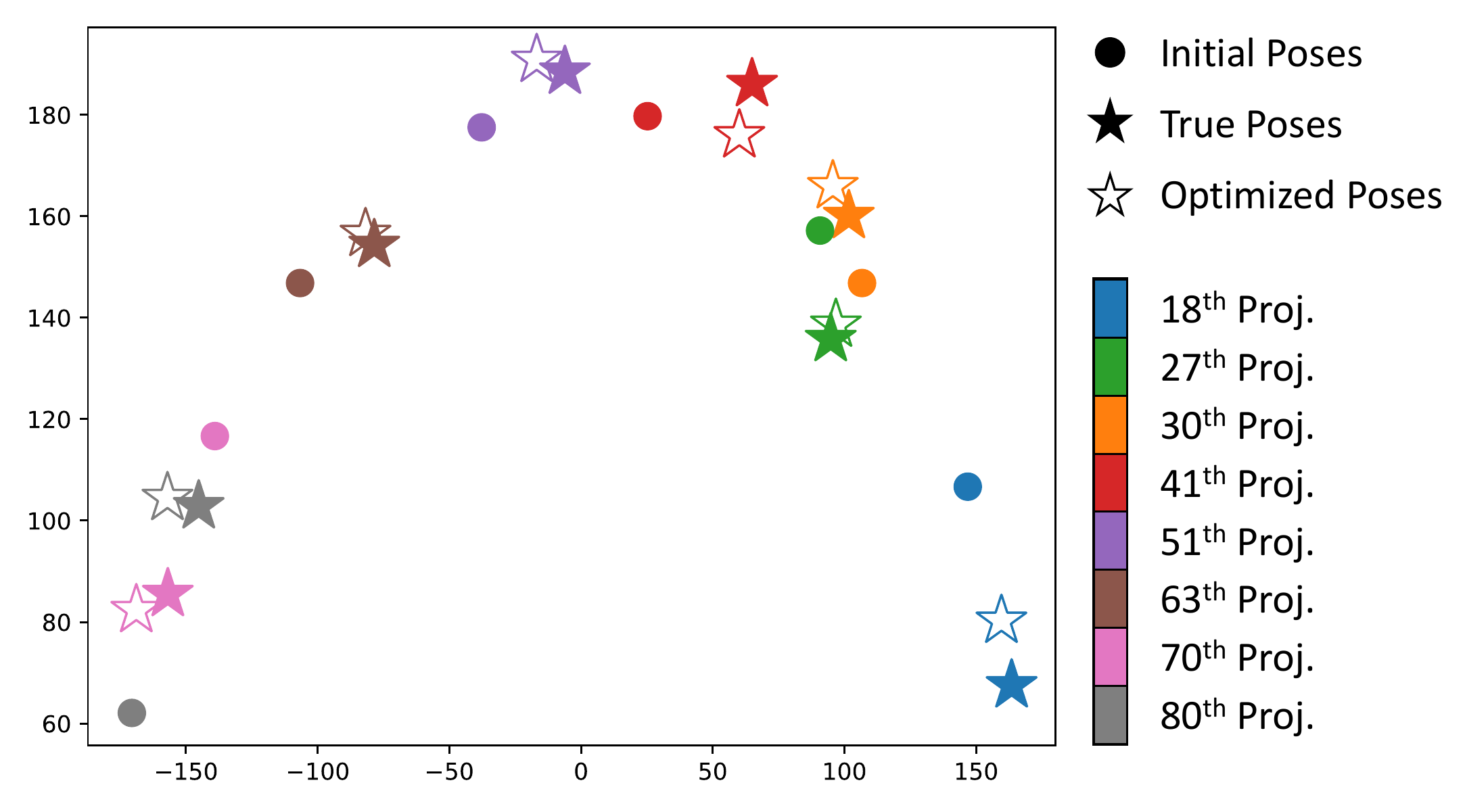}
    \caption{Visual comparison of initial, true, and optimized projection poses on a sample of the COVID-19 dataset \cite{shakouri2021covid19} for SVCT with severe rigid motion ($\pm16^\circ\&\pm16\ \mathsf{pixels}$).}
    \label{fig:pose}
\end{figure}

 We conduct an ablation study to confirm the effectiveness of the pose correction in our model. Table \ref{tab:my_label} shows the qualitative results. We can observe that the pose correction significantly improves the model performance. For example, PSNR and SSIM respectively improve $4.39$ dB ($24.17$ vs $19.78$) and $0.3198$ ($0.8694$ vs $0.5496$) for the mild rigid motion ($\pm2^\circ\&\pm2\ \mathsf{pixels}$). Fig. \ref{fig:pose_com} shows the qualitative results for the severe rigid motion ($\pm16^\circ\&\pm16\ \mathsf{pixels}$). Obviously, the proposed model with the pose correction could recover image structure and details clearly while the model without the pose correction performs very poorly. We also demonstrate the visual comparison of the initial (i.e., inaccurate), true, and optimized poses in Fig. \ref{fig:pose}. We can observe that all the projection poses are almost precisely corrected. Overall, the pose correction module is crucial for our model performance.
\begin{table}[t]
    \centering
    \resizebox{0.48\textwidth}{!}{  
    \begin{tabular}{c|c|c}
        \textbf{Motion Settings} &  \textbf{w/o Pos. Corr.} & \textbf{w/ Pos. Corr.}\\ \hline
        $\pm2^\circ\&\pm2\ \mathsf{pixels}$ & $19.78/0.5496/0.3060$ & $24.17/0.8694/0.0276$\\
        $\pm8^\circ\&\pm8\ \mathsf{pixels}$ & $13.71/0.2860/0.4708$ & $19.60/0.7564/0.0387$\\
        $\pm16^\circ\&\pm16\ \mathsf{pixels}$ & $10.58/0.2071/0.5794$ & $16.85/0.6339/0.0570$\\ \hline
        \textbf{Mean} & $14.69/0.3476/0.4521$ & $20.21/0.7532/0.0411$\\
    \end{tabular}}
    \caption{Quantitative results (PSNR/SSIM/LPIPS) of our model without and with poses correction on the COVID-19 dataset \cite{shakouri2021covid19} for SVCT with three levels of rigid motion.}
    \label{tab:my_label}
\end{table}

\section{Conclusion and Limitation}
\label{sec:conc}
\par This work proposes a novel self-supervised NeRF-based model for SVCT reconstruction. Different from existing NeRF-based SVCT methods, the proposed model extra models the rigid motion in the CT acquisition process. Therefore, it can reconstruct robust and high-quality CT results from the rigid motion-corrupted measurements. Experiments on a public CT dataset indicate that our proposed model are greatly superior to two latest NeRF-based method for SVCT reconstruction with rigid motion.
\par Although the proposed model obtains excellent reconstruction performance for SVCT imaging task with rigid motion, there still are two limitations: (1) Our model is based on the 2D parallel X-ray beam CT, while more advanced types of X-ray beams (\eg 2D fan beam, 3D cone beam) are not implemented; (2) Our model can calibrate the rigid motion in the CT acquisition, while more complex non-rigid motion cannot be handled now.

\section{Compliance with ethical standards}
\label{sec:ethics}
\par This research study is conducted retrospectively using human subject data made available in open access by Harvard Dataverse, COVID19-CT Dataset \cite{shakouri2021covid19} (https://doi.org/10.7910/DVN/6ACUZJ). Ethical approval was not required as confirmed by the license attached with the open-access data.
\section{Acknowledgments}
\label{sec:acknowledgments}
\par This work is supported by the National Natural Science Foundation of China (No. 62071299, 61901256, 91949120).
\bibliographystyle{IEEEbib}
\bibliography{refs}

\end{document}